\documentclass[11pt,pdftex, preprint]{elsarticle}
\usepackage[usenames,dvipsnames,svgnames,table]{xcolor}
\usepackage[colorlinks]{hyperref}
\AtBeginDocument{\hypersetup{linkcolor=blue, citecolor=blue}}
\usepackage[T1]{fontenc}
\usepackage[font=footnotesize,labelfont=bf]{caption}
\usepackage{bm, bbm, bbding}
\usepackage{seqsplit}
\usepackage{amssymb,amsmath,amsfonts,mathrsfs,graphicx,caption,kantlipsum,units,yfonts}
\usepackage{txfonts,dsfont}
\usepackage{feynmf}

\usepackage{pdflscape}
\usepackage{rotating}
\usepackage[toc,page,title,titletoc,header]{appendix}
\usepackage{lineno,setspace, framed, hyperref}
\usepackage[top=2.6cm, bottom=2.7cm, left=2.8cm, right=2.8cm]{geometry}
\usepackage[novbox]{pdfsync}
\usepackage{subcaption}

\usepackage[numbers]{natbib}

\makeatletter
\def\ps@pprintTitle{%
  \let\@oddhead\@empty
  \let\@evenhead\@empty
  \def\@oddfoot{\reset@font\hfil\thepage\hfil}
  \let\@evenfoot\@oddfoot
}
\makeatother

\journal{}
\begin{document}

\begin{frontmatter}

\title{Big Bang nucleosynthesis and baryogenesis in power-law $f(R)$ gravity: Revised constraints from the semianalytical approach}
\author{David Wenjie Tian\fnref{myfootnote}}
\address{Faculty of Science,  Memorial University, St. John's, Newfoundland, A1C 5S7, Canada}
\fntext[myfootnote]{Email address: wtian@mun.ca}

\begin{abstract}
In this paper we investigate the primordial nucleosynthesis in $\mathscr{L}=\varepsilon^{2-2\beta}R^\beta+{16\pi}m_P^{-2}\mathscr{L}_m$ gravity, where $\varepsilon$ is a constant balancing the dimension of the field equation, and $1<\beta<(4+\sqrt{6})/5$ for the positivity of energy density and temperature. From the semianalytical approach, the influences of $\beta$ to the decoupling of neutrinos, the freeze-out temperature and concentration of nucleons, the opening of deuterium bottleneck, and the $^4$He abundance are all extensively analyzed; then $\beta$ is constrained to $1<\beta<1.05$ for $\varepsilon=1$ [1/s] and $1<\beta<1.001$ for $\varepsilon=m_P$ (Planck mass). 
Supplementarily from the empirical approach, abundances of the lightest elements (D, $^4$He, $^7$Li) are computed by the model-independent best-fit formulae for nonstandard primordial nucleosynthesis, and we find the constraint $1< \beta \leq 1.0505$ which corresponds to the extra number of neutrino species $0< \Delta N_\nu^{\text{eff}} \leq 0.6365$; also, the $^7$Li abundance problem cannot be solved by $\mathscr{L}=\varepsilon^{2-2\beta}R^\beta+{16\pi}m_P^{-2}\mathscr{L}_m$ gravity for this domains of $\beta$. Finally, the consistency with the mechanism of gravitational baryogenesis is estimated. \\

\end{abstract}

\end{frontmatter}


\section{Introduction}

In the past few decades, the increasingly precise measurements for the cosmic abundances of the lightest elements have imposed stringent constraints to the thermal history of the very early Universe. The observed protium, deuterium (D) and $^4$He abundances prove to agree well with those predicted by the standard Big Bang nucleosynthesis (BBN) in general relativity (GR).

As is well known, any
modification to the Hubble expansion rate and the time-temperature correspondence would affect the decoupling of neutrinos, the freeze-out of nucleons, the time elapsed to open the deuterium bottleneck, and the abundances of $^4$He along with the other light elements. To better meet the observations from the very early Universe,
nonstandard BBN beyond the SU$(3)_c\times$SU$(2)_W\times$U$(1)_Y$ minimal standard model \cite{Nonstandard BBN} or beyond the standard gravitational framework of GR have also received  a lot of discussion, such as nonstandard BBN in scalar-tensor gravity \cite{Nonstandard BBN scalar-tensor, Nonstandard BBN scalar-tensor II, Nonstandard BBN scalar-tensor III, Nonstandard BBN scalar-tensor IV}, Brans-Dicke gravity with a varying energy term related to the cosmic radiation background \cite{Nonstandard BBN Brans-Dicke Lambda, Nonstandard BBN Brans-Dicke II}, $f(R)$ gravity \cite{BBN fR I, BBN fR II, BBN fR III},  $f(\mathcal{G})$ generalized Gauss-Bonnet gravity \cite{Nonstandard BBN Gauss Bonnet}.
Nonstandard BBN in helps constrain these modified gravities from the properties of the very early Universe, which supplements the popular constraints from the accelerated expansion of the late-time Universe.

So far, nonstandard BBN in $f(R)\propto R^\beta$ gravity has been involved in Ref.\cite{BBN fR I} and studied in Refs.\cite{BBN fR II, BBN fR III}; however, these earlier investigations are not satisfactory. Ref.\cite{BBN fR I} only calculated the nonstandard decoupling temperature of nucleons; the BBN energy scale was inappropriately extended to $T\leq 100$ MeV, and the interconversion rate $\Gamma_{n\to p}$ between neutrons and protons was incorrectly equated to the approximate rate at the high-energy domain $T\gg m_n-m_p\simeq 1.2933$ MeV. Ref.\cite{BBN fR II} continued to investigate the primordial $^4$He synthesis  in $f(R)\propto R^\beta$ gravity from a semianalytical approach; however, the BBN process after neutrinos' decoupling was numerically calculated using the standard Hubble expansion of GR rather than the generalized Hubble rate
in $f(R)\propto R^\beta$ gravity. Also, due to the inconsistent setups of the geometric conventions, the domain of $\beta$ was incorrectly set as $(4-\sqrt{6})/5< \beta<1$ in Refs.\cite{BBN fR I, BBN fR II}, which had led to quite abnormal behaviors for $\beta\approx (4-\sqrt{6})/5$.  Ref.\cite{BBN fR III} corrected the domain of $\beta$ into $1< \beta<(4+\sqrt{6})/5$, and re-constrained the parameter $\beta$ by the abundances of both deuterium and $^4$He; however, the computations were carried out using the public BBN code, and the details regarding the influences of $\beta$ to the BBN process were not brought to light.

In this work, we aim to overcome the flaws in Refs.\cite{BBN fR II,BBN fR III}, and reveal every detail of the BBN process in $f(R)\propto R^\beta$ gravity. This paper is analyzed as follows. Section~\ref{Sec Generalized Friedmann equations in fR gravity} introduces the generalized Friedmann equations for the radiation-dominated Universe in generic $f(R)$ gravity. In Sec.~\ref{Sec Power-law fR gravity}, the power-law $f(R)$ gravity with the total Lagrangian density $\mathscr{L}=\varepsilon^{2-2\beta}R^\beta+{16\pi}m_P^{-2}\mathscr{L}_m$ is set up ($\varepsilon$ being some constant balancing the dimensions of the field equation), with the nonstandard Hubble expansion and the generalized time-temperature relation derived. The decoupling of neutrinos is studied in Sec.~\ref{Sec equilibrium of nucleons and weak freeze out of neutrinos}, while the temperature and neutrons' concentration at the nucleon freez-out are computed in Sec.~\ref{Sec Temperature and concentrtion at nucleon freeze out}.
In Sec.~\ref{Sec Opening of deuterium bottleneck and helium synthesis}, the opening of the deuterium bottleneck and the primordial $^4$He abundance are found out, which exerts constraints to the parameter $\beta$ compared with the $^4$He abundance in astronomical measurement. The semianalytical discussion in Secs.~\ref{Sec equilibrium of nucleons and weak freeze out of neutrinos}$\sim$\ref{Sec Opening of deuterium bottleneck and helium synthesis} for $\mathscr{L}=\varepsilon^{2-2\beta}R^\beta+{16\pi}m_P^{-2}\mathscr{L}_m$ gravity is taken the GR limit $\beta\to 1$ in Sec.~\ref{Sec GR limit} to recover the standard BBN.  Moreover, the primordial abundances of deuterium, $^4$He and $^7$Li are calculated in Sec.~\ref{Sec Empirical constraints from D and 4He abundances} from the empirical approach using the model-independent best-fit formulae, which supplements the results from the semianalytical approach. Finlly, the consistency of $\mathscr{L}=\varepsilon^{2-2\beta}R^\beta+{16\pi}m_P^{-2}\mathscr{L}_m$ gravity with the gravitational baryogenesis is estimated in Sec.~\ref{Sec Consistency with gravitational baryogenesis}.

Throughout this paper, for the physical quantities involved in the thermal history of the early Universe, we use the natural unit system of particle physics which sets $c=\hbar=k_B=1$ and is related with le syst\`eme international d'unit\'es by $1\text{ MeV}=1.16\times 10^{10}\text{ kelvin} = 1.78\times 10^{-30}\text{ kg}=(1.97\times 10^{-13} \text{ meters})^{-1}= (6.58\times 10^{-22}\text{ seconds})^{-1}$. On the other hand, for the spacetime geometry, we adopt the conventions $\Gamma^\alpha_{\beta\gamma}=\Gamma^\alpha_{\;\;\,\beta\gamma}$,
$R^{\alpha}_{\;\;\beta\gamma\delta}=\partial_\gamma \Gamma^\alpha_{\delta\beta}\cdots$ and $R_{\mu\nu}=R^\alpha_{\;\;\mu\alpha\nu}$ with the metric signature $(-,+++)$.


\section{Generalized Friedmann equations in $f(R)$ gravity}
\label{Sec Generalized Friedmann equations in fR gravity}

As a straightforward generalization of the Hilbert-Einstein action $S_{\text{HE}}=\int \!\sqrt{-g}\,d^4x \left(R+{16\pi}m_P^{-2}\mathscr{L}_m\right)$, $f(R)$ gravity is given by the action
\begin{equation}\label{f(R) action}
\mathcal{S}=\int d^4x\!\sqrt{-g}\,\left[f(R,\varepsilon)+{16\pi}m_P^{-2} \mathscr{L}_m\right]\,,
\end{equation}
where $R$ is the Ricci scalar of the spacetime, and $\varepsilon$ is some constant balancing the dimensions of the field equation.
Also, $m_P$ refers to the Plank mass, which is related to Newton's constant $G$ by
$m_P\coloneqq 1/\!\sqrt{G}$ and takes the value $m_P\simeq 1.2209\times 10^{22}\text{ MeV}$. 
Variation of Eq.(\ref{f(R) action})
with respect to the inverse metric $\delta S/\delta g^{\mu\nu}=0$ yields the field equation
\begin{equation}\label{Field Eq generic fR}
f_R R_{\mu\nu}-\frac{1}{2}f+\left(g_{\mu\nu}\Box-\nabla_\mu\nabla_\nu\right)f_R={8\pi}m_P^{-2} \mathcal{T}_{\mu\nu}^{(m)}\,,
\end{equation}
where $f_R\coloneqq df(R,\varepsilon)/dR$, $\Box$ denotes the covariant d'Alembertian $\Box\coloneqq g^{\alpha\beta}\nabla_\alpha\nabla_\beta$,  and the stress-energy-momentum tensor $\mathcal{T}_{\mu\nu}^{(m)}$ of the physical content is defined by the matter Lagrangian density $\mathscr{L}_m$ via $\mathcal{T}_{\mu\nu}^{(m)}\coloneqq \frac{-2}{\!\sqrt{-g}}\frac{\delta\left(\sqrt{-g}\mathscr{L}_m\right)}{\delta g^{\mu\nu}}$. This paper considers the spatially flat, homogeneous and isotropic Universe, which, in the $(t,r,\theta,\varphi)$ comoving coordinates along the cosmic Hubble flow, is depicted by the Friedmann-Robertson-Walker (FRW) line element
\begin{equation}\label{FRW metric}
\begin{split}
ds^2 = -dt^2+ a(t)^2 dr^2 + a(t)^2 r^2
 \Big( d\theta^2+\sin^2 \!\theta d\varphi^2 \Big)\,,
\end{split}
\end{equation}
where $a(t)$ denotes the cosmic scale factor. Assume a perfect-fluid material content $\mathcal{T}^{\mu(m)}_{\;\;\nu} = \text{diag} [-\rho, P, P,$ $P]$, with $\rho$ and $P$ being the energy density and pressure, respectively. Then Eq.(\ref{Field Eq generic fR}) under the flat FRW metric yields the generalized Friedmann equations
\begin{equation}\label{Friedmann EqI generic FR}
3\frac{\ddot{a}}{a}f_R-\frac{1}{2}f
-3\frac{\dot{a}}{a}f_{RR}\dot{R}=-{8\pi}m_P^{-2} \rho\,,
\end{equation}
\begin{equation}\label{Friedmann EqII generic FR}
\left(\frac{\ddot{a}}{a}+2\frac{\dot{a}^2}{a^2} \right)f_R-\frac{1}{2}f- f_{RR}\ddot{R}
- f_{RRR}(\dot{R})^2-3\frac{\dot{a}}{a}f_{RR}\dot{R}= {8\pi}m_P^{-2} P\,,
\end{equation}
where overdot denotes the derivative with respect to the comoving time, $f_{RR}\coloneqq d^2f(R,\varepsilon)/dR^2$, and $f_{RRR}\coloneqq d^3f(R,\varepsilon)/dR^3$. Moreover, the equation of local energy-momentum conservation, $\nabla^\mu \mathcal{T}_{\mu\nu}^{(m)}=0$, gives rise to
\begin{equation}\label{Continuity Eq}
\dot{\rho}+3\frac{\dot{a}}{a}
(\rho+P)=0\,.
\end{equation}
When $f(R,\epsilon)=R$, one recovers Einstein's equation $R_{\mu\nu}-\frac{1}{2}Rg_{\mu\nu}= {8\pi}m_P^{-2}  \mathcal{T}_{\mu\nu}^{(m)}$ for GR, as well as the standard Friedmann equations $\displaystyle 3\frac{\dot{a}^2}{a^2}=-{8\pi}m_P^{-2} \rho$ and $\displaystyle 3\frac{\ddot a}{a}=-{4\pi}m_P^{-2}(\rho+P)$.

The very early (i.e. the first few minutes) Universe is absolutely radiation-dominated, with the equation of state $\rho=3P$.
Thus Eq.(\ref{Continuity Eq}) integrates and yields that the radiation density is related to the cosmic scale by
\begin{equation}
\rho\,=\,\rho_0 a^{-4}\,\propto\,a^{-4}\,.
\end{equation}
$\rho$ attributes to the energy densities of all relativistic species, which are exponentially greater than those of the nonrelativistic particles, and
therefore $\rho=\sum \rho_i(\text{boson})+\frac{7}{8}\sum \rho_j(\text{fermion})=\sum \frac{\pi^2}{30}g_i^{(b)} T_i^4(\text{boson})+\frac{7}{8}\sum \frac{\pi^2}{30}g_j^{(f)} T_j^4(\text{fermion})$, where $\{g_i^{(b)}, g_j^{(f)}\}$ are the numbers of statistical degrees of freedom for relativistic bosons and fermions, respectively. More concisely, normalizing the temperatures of all relativistic species with respect to photons' temperature $T_\gamma \equiv T$, one has the generalized Stefan-Boltzmann law
\begin{equation}\label{rho and g for radiation}
\rho=\frac{\pi^2}{30}g_*T^4\qquad\text{with}\qquad g_*\coloneqq\sum_{\text{boson}} g_i^{(b)}
\left(\frac{T_{i}}{T}\right)^4
+\frac{7}{8}\sum_\text{fermion} g_j^{(f)} 
\left(\frac{T_{j}}{T}\right)^4\,,
\end{equation}
where, in thermodynamic equilibrium, $T$ is the common temperature of all relativistic particles.


\section{Power-law $f(R)$ gravity}\label{Sec Power-law fR gravity}

This paper works with the specific power-law nonlinear gravity 
\begin{equation}\label{Power law fR action}
\mathcal{S}
=\int d^4x\!\sqrt{-g}\,\left(\varepsilon^{2-2\beta}R^\beta+{16\pi}m_P^{-2}\mathscr{L}_m\right)\,,
\end{equation}%
where $\beta=\text{constant}>0$.
Recall that for GR with $\beta=1$, the first Friedmann equation $3\dot{a}^2/a^2=-{8\pi}m_P^{-2} \rho_0 a^{-4}$ yields the behavior $a=a_0 t^{1/2}\propto  t^{1/2}$. Similarly, assume a power-law solution ansatz $a=a_0 t^\alpha$ with the index $\alpha=\text{constant}>0$ -- note that this ansatz proves valid for $\mathscr{L}=\varepsilon^{2-2\beta}R^\beta+{16\pi}m_P^{-2}\mathscr{L}_m$ gravity, though  invalid for generic $f(R)$ gravity. This way, the generalized first Friedmann equation (\ref{Friedmann EqI generic FR}) yield 
\begin{equation}
\beta=2\alpha\quad,\quad H\coloneqq\frac{\dot a}{a}=\frac{\beta}{2t}\,,
\end{equation}
and
\begin{equation}\label{Concrete Friedmann EqI generic FR}
\left[\frac{12(\beta-1)}{\beta}H^2\right]^\beta \frac{\big(-5\beta^2+8\beta-2\big)}{\beta-1}=32\pi \varepsilon^{2\beta-2}m_P^{-2}\rho\,, 
\end{equation}
where $H$ refers to the cosmic Hubble parameter.
The weak, strong and dominant energy conditions for classical matter fields require the energy density $\rho$ to be positive definite, and as a consequence, the positivity of the left hand side of Eq.(\ref{Concrete Friedmann EqI generic FR}) limits $\beta$ to the domain
\begin{equation}\label{beta domain I}
1<\beta<\frac{4+\sqrt{6}}{5}\simeq 1.2899\,.
\end{equation}
Note that the Ricci scalar for the flat FRW metric with $a=a_0 t^{\beta/2}$ reads
\begin{equation}
R=6\left(\frac{\ddot{a}}{a}+\frac{\dot{a}^2}{a^2}\right)=\frac{3\beta(\beta-1)}{t^2} \,,
\end{equation}
so $R>0$ and $R^\beta$ is always well defined in this domain.

Eqs.(\ref{rho and g for radiation}) and (\ref{Concrete Friedmann EqI generic FR}) imply that the expansion rate of the Universe is related to the radiation temperature by
\begin{equation}\label{Hubble evolution}
\begin{split}
H
&= \left(\frac{\beta}{12(\beta-1)}\right)^{1/2}\left(\frac{(\beta-1)\,g_*}{-5\beta^2+8\beta-2}\right)^{1/2\beta}
\left(\sqrt{\frac{32\pi^3}{30}}\frac{T^2}{m_P}\right)^{1/\beta}
\varepsilon^{1-\frac{1}{\beta}}\\
&= 0.2887\times\sqrt{\frac{\beta}{\beta-1}}\times\left(\sqrt{\frac{(\beta-1)\,g_*}
{-5\beta^2+8\beta-2}}\right)^{1/\beta}
\left(0.7164\cdot T^2_{\text{MeV}}\right)^{1/\beta} \varepsilon_{s}^{1 -1/\beta}\text{ [s}^{-1}]\,,
\end{split}
\end{equation}
where $T_{\text{MeV}}$ refers to the value (dimensionless) of temperature evaluated in the unit of MeV, $T=T_{\text{MeV}}\times\text{ [1\,MeV]}$, $\varepsilon_{s}$ is the value of $\varepsilon$ in the unit of [1/s], and numerically ${T^2}/{m_P}={T^2_{\text{MeV}}}/{8.0276}\text{ [s}^{-1}]$.

Moreover, as time elapses after the Big Bang,
the space expands and the Universe cools. Eq.(\ref{Hubble evolution}) along with $H=\beta/(2t)$ lead to the time-temperature relation
\begin{equation}\label{Time-temperature relation}
\begin{split}
t
&=\sqrt{3\beta(\beta-1 )}\, \left(\sqrt{\frac{ -5\beta^2+8\beta-2 }{ (\beta-1)\,g_*}}\right)^{1/\beta}
\left(\sqrt{\frac{30}{32\pi^3}}\frac{m_P}{T^2}\right)^{1/\beta} \varepsilon^{1/\beta -1}\\
&=\sqrt{3\beta(\beta-1 )}\, \left(\sqrt{\frac{ -5\beta^2+8\beta-2 }{ (\beta-1)\,g_*}}\right)^{1/\beta}
\left(\frac{1.3959}{T^2_{\text{MeV}}}\right)^{1/\beta} \varepsilon_s^{1/\beta -1}\text{ [s]} \,.
\end{split}
\end{equation}
Eqs.(\ref{Hubble evolution}) and (\ref{Time-temperature relation}) play important roles in studying the primordial nucleosynthesis and the baryogenesis. For the calculations in the subsequent sections, we will utilize two choices of $\varepsilon$ to balance the dimensions:
\begin{enumerate}
  \item[(a)]  $\varepsilon=1 \text{ [s}^{-1}]$. This choice can best respect and preserve existent investigations in mathematical relativity for the $f(R)$ class of modified gravity, which have been analyzed for $\mathscr{L}=f(R)+16\pi m_P^{-2}\mathscr{L}_m$ without caring the physical dimensions. Supplementarily,
      we have $\varepsilon_s=6.58\times 10^{-22}$ MeV.

  \item[(b)]  $\varepsilon=m_P$. The advantage of this choice is there is no deed to employ new parameters outside the mathematical expression $\mathscr{L}=f(R)+16\pi m_P^{-2}\mathscr{L}_m$. Supplementarily, $m_{ps}\coloneqq m_P\text{[1/s]} \simeq 0.1854\times 10^{44}\text{ [s}^{-1}]$.
      However, the disadvantage of this choice is also apparent. For example,  Eq.(\ref{Power law fR action}) with $\varepsilon=m_P$ is mathematically and gravitationally equivalent to
      \begin{equation}\label{Power law fR criticism}
       \mathcal{S}
       =\int d^4x\!\sqrt{-g}\,\left(R^\beta+{16\pi}m_P^{2\beta-4}\mathscr{L}_m\right)
       =\int d^4x\!\sqrt{-g}\,\left(R^\beta+{16\pi}G^{2-\beta}\mathscr{L}_m\right)\,,
       \end{equation}
       and thus hence, the deviation between $f(R)=R$ and $f(R)=R^\beta$ would indicate a departure of the matter-gravity coupling strength from Newton's contant $G$ to $G^{2-\beta}$; as a consequence, to constrain the parameter $\beta$ for $\mathscr{L}=R^\beta/m_P^{2\beta-2}+{16\pi}m_P^{-2}\mathscr{L}_m$ gravity, one just need to check the measurement of Newton's constant rather than recalculate the testable gravitational processes.
\end{enumerate}

\section{Chemical and thermal equilibrium of nucleons, and weak freeze-out of neutrinos}\label{Sec equilibrium of nucleons and weak freeze out of neutrinos}

According to the SU$(3)_c\times$SU$(2)_W\times$U$(1)_Y$ minimal standard model,
primordial nucleosynthesis happens after the temperature drops below $T=10$ MeV, when all mesons have decayed into nucleons. At $T\approx 10$ MeV, photons are in thermal equilibrium with neutrons and protons, which are interconverted by the two-body reactions
\begin{equation}\label{baryon interconversions I}
n+\nu_e\rightleftharpoons p+e^-\;,\quad
n+e^+\rightleftharpoons p+\bar\nu_e\,,
\end{equation}
as well as the neutron decay/fusion
\begin{equation}\label{baryon interconversions II}
n\rightleftharpoons 
p+e^-+\bar\nu_e\,.
\end{equation}
When the nuclear reaction rate $\Gamma(n\rightleftharpoons p)$ is faster than the Hubble expansion rate, the interconversions Eqs.(\ref{baryon interconversions I}) and (\ref{baryon interconversions II}) are fast enough to maintain neutrons and protons in thermal equilibrium.

Introduce the following dimensionless quantity for the number concentration of neutrons among all baryons,
\begin{equation}
X_n=\frac{n_n}
{n_n+n_p}\,,
\end{equation}
and thus before the opening of the deuterium bottleneck, the proton concentration is $X_p=1-X_n=\frac{n_p}{n_n+n_p}$. Regard neutrons and protons as the two energy states of nucleons, and the approximated Maxwell-Boltzmann energy distribution function yields
\begin{equation}
\frac{X_n^{\text{eq}}}{X_p^{\text{eq}}}=\frac{n_n}{n_p}
=\exp\left(-\frac{ Q }{T}+\frac{\mu_e-\mu_{\nu_e}}{T}\right)
\simeq \exp\left(-\frac{ Q }{T}\right)\,,
\end{equation}
or
\begin{equation}\label{Xn equilibrium}
X_n^{\text{eq}}=\frac{1}{1+
\exp\left(\frac{ Q }{T}\right)}\,,
\end{equation}
where $ Q \coloneqq m_n-m_p=1.2933 \text{ MeV}$ denotes the neutron-proton mass difference (with $m_n=939.5654$ MeV, $m_P=938.2721$ MeV),
and we have applied the standard-model assumption $\mu_{\nu_e}=0$ and the fact that $\mu_e\ll T$ for $T\gtrsim 0.03$ MeV.
Eq.(\ref{Xn equilibrium}) implies that $X_n^{\text{eq}}\to 1/2=X_p^{\text{eq}}$ for $T\gg 1.2933 \text{ MeV}$, and $X_n^{\text{eq}}$ gradually decreases as the temperature drops, while nucleons remain in weak-interaction equilibrium until neutrinos decouple.



\begin{tabular}{c}

\begin{minipage}{\linewidth}
\captionof{figure}{$T_f^\nu$ (in MeV) for $\varepsilon=1 \text{ sec}^{-1}=6.58\times 10^{-22}$ MeV}\label{FigWeakFreezeTemI}\vspace{-1.25cm}
 \makebox[1\linewidth]{ \includegraphics[keepaspectratio=true,scale=0.58]{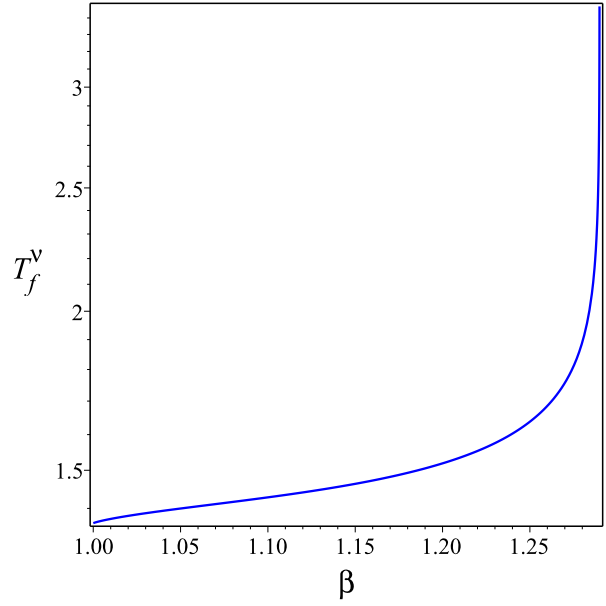}}
\end{minipage}
\vspace{-6.5cm}  \\

 \begin{minipage}{\linewidth}
\captionof{figure}{$T_f^\nu$ (in MeV) for $\varepsilon=m_P=1.2209\times 10^{-22}$ MeV}\label{FigWeakFreezeTemII}\vspace{-1.25cm}
 \makebox[1\linewidth]{ \includegraphics[keepaspectratio=true,scale=0.58]{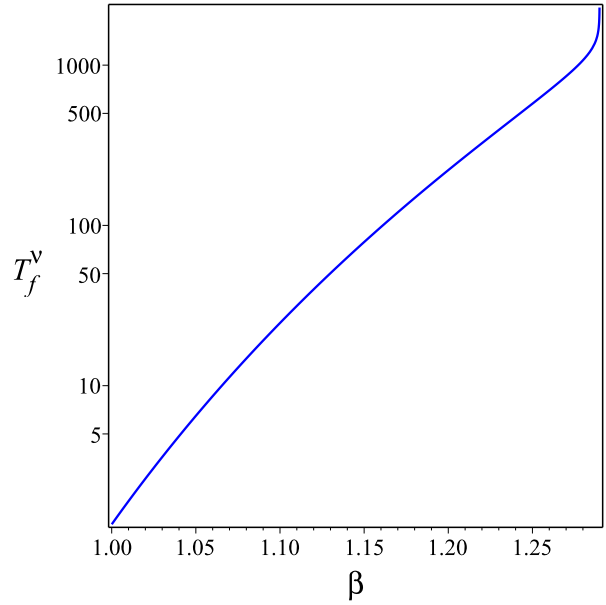}}
\end{minipage}
\vspace{-6.5cm}

\end{tabular}

Neutrinos are in equilibrium with photons, nucleons and electrons via weak interactions   and elastic scattering. The interaction rate is
\begin{equation}\label{Neutrino interaction rate}
\Gamma_{\nu_e}
\simeq 1.3 G_F^2 T^5 \simeq 0.2688\, T^5_{\text{MeV}} \text{ [s}^{-1}]\,,
\end{equation}
where
$G_F$ is Fermi's constant in beta decay and generic weak interactions, and $G_F=1.1664\times 10^{-11} \text{MeV}^{-2}$.  Neutrinos decouple when $\Gamma_{\nu_e}=H$, and according to Eqs.(\ref{Hubble evolution}) and (\ref{Neutrino interaction rate}), the weak freeze-out temperature $T_f^\nu$ is the solution to
\begin{equation}\label{Neutrino decoupling temperature}
T^{5-2/\beta}_{\text{MeV}} = 1.0741\,\times\sqrt{\frac{\beta}{\beta-1}}\times\left(0.7164\cdot\sqrt{\frac{(\beta-1)\,g_*}
{-5\beta^2+8\beta-2}}\right)^{1/\beta}
\varepsilon_{s}^{1 -1/\beta}\,.
\end{equation}


Fig.\ref{FigWeakFreezeTemI} and Fig.\ref{FigWeakFreezeTemII} illustrate the dependence of $T_f^\nu$ on $\beta$ for two different choices of $\varepsilon$, and some typical values of $T_f^\nu$ have been collected in  Tables \ref{Table I} and \ref{Table II}.
Note that in the calculation of $T_f^\nu$, we have used $g_*=g_*(T: 1\sim10 \text{ MeV})$, $g_b=2\text{ (photon)}$ and $g_f=2\times2 \,(e^{\pm})+ 2\times3.046\text{ (neutrino)}=10.092$, and thus the effective number of degree of freedom $g_*=g_b+\frac{7}{8}g_f=10.8305$, with all these relativistic species in thermal equilibrium at the same temperature. That is to say, the effective number of species for light neutrinos is set to be $N_{\text{eff}}=3.046$ rather than  $N_{\text{eff}}=3$; this correction attributes to the fact that the neutrino decoupling is actually a thermal process of finite time rather than an instantaneous event \cite{Efftive neutrino species 3.046}.\\

\section{Temperature and concentrtion at nucleon freeze-out}\label{Sec Temperature and concentrtion at nucleon freeze out}

\subsection{Temperature for freeze-out of nucleons}\vspace{2mm}

After the weak freeze-out of neutrinos, the neutron concentration deviates from the equilibrium value in Eq.(\ref{Xn equilibrium}), and the evolution of $X_n$ satisfies
\begin{equation}\label{Xn Nonequilibrium}
\begin{split}
\frac{dX_n}{dt}=-\Gamma_{n\to p}X_n
+\Gamma_{p\to n}\left(1-X_n\right)
=-\Gamma_{n\to p}\left(1+e^{-\frac{ Q }{T}} \right) 
\left(X_n -X_n^{\text{eq}}\right)\,,
\end{split}
\end{equation}
where $\Gamma_{n\to p}$/$\Gamma_{p\to n}$ denotes the reaction rate to convert neutrons/protons into protons/neutrons.
When nucleons and leptons are carried apart by the Hubble expansion faster than their collisions, the reactions in Eqs.(\ref{baryon interconversions I}) and (\ref{baryon interconversions II}) cease and $X_n$ freezes out.

On the other hand, as shown in Fig.\ref{FigWeakFreezeTemI} and Fig.\ref{FigWeakFreezeTemII}, $T^\nu_f$ is positively related to $\beta$, and $T^\nu_f$ is always minimized in the GR limit $\beta\to1$, with
$\displaystyle\min(T^\nu_f)=\lim_{\beta\to1} T^\nu_f$=1.3630 MeV; moreover, Eq.(\ref{Time-temperature relation}) indicates $t\propto T^{-2/\beta}$, thus by setting $T=\min(T^\nu_f)$=1.3630 MeV in Eq.(\ref{Time-temperature relation}) one gets the upper limit of $t_f^\nu$ for any $1<\beta<(4+\sqrt{6})/5$, which is depicted in Fig.\ref{FigNeuDecoupTimeI} and Fig.\ref{FigNeuDecoupTimeII} with $\tau_f^\nu\coloneqq\max(t_f^\nu)$. Thus, the mean time of neutron decay $\tau_n=880.0\pm 0.9$ [s] \cite{BBN Particle Data Group} is always far greater than the time elapsed from big bang to neutrino freeze-out. Furthermore, as will be shown in Tables \ref{Table I} and \ref{Table II}, the time $t_f^n$ by nucleons' freeze-out also happens within the first few seconds and is still far less than $\tau_n$.
Hence, the rate of three-body reaction in Eq.(\ref{baryon interconversions II}) is negligible by the stage of free-out.

The combined reaction rate $\Gamma_{n\to p}$ for the two-body reactions in Eq.(\ref{baryon interconversions I}) is \cite{BBN helium simplified}
\begin{equation}\label{Combined two-body reaction rate}
\Gamma_{n\to p}=\frac{255}{\tau_n}\left(\frac{T}{ Q }\right)^5
\left[\left(\frac{ Q }{T}\right)^2+6\left(\frac{ Q }{T}\right)+12  \right]  \text{ [s}^{-1}]\,.
\end{equation}
Introduce the dimensionless variable
\begin{equation}
x\coloneqq \frac{ Q }{T}\,, 
\end{equation}
and then Eq.(\ref{Combined two-body reaction rate}) becomes $\Gamma_{n\to p}=\frac{255}{\tau_n}x^{-5}(x^2+6x+12) \text{ [s}^{-1}]$. Also, the Hubble parameter can be recast into\\

\begin{tabular}{c}

\begin{minipage}{\linewidth}
\captionof{figure}{$\tau_f^\nu=\max\;t_f^\nu$ (in [sec]) for $\varepsilon=1 \text{ sec}^{-1}=6.58\times 10^{-22}$ MeV, T=1.3630 MeV}\label{FigNeuDecoupTimeI}\vspace{-1.25cm}
 \makebox[1\linewidth]{ \includegraphics[keepaspectratio=true,scale=0.58]{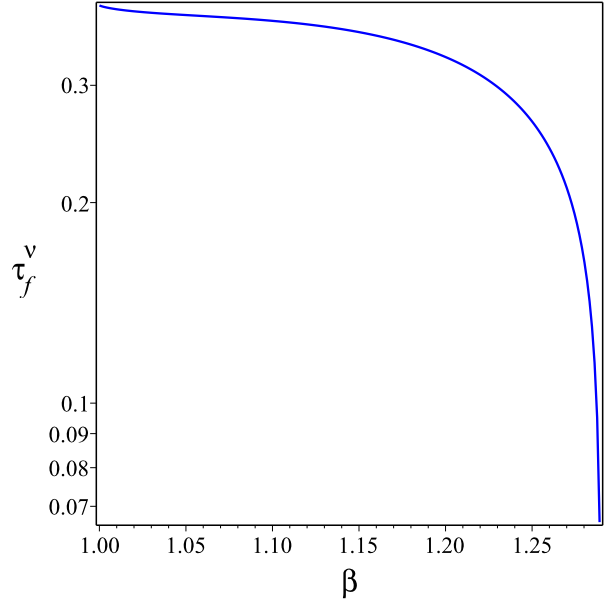}}
\end{minipage}
\vspace{-6.5cm}  \\

 \begin{minipage}{\linewidth}
\captionof{figure}{$\tau_f^\nu=\max\;t_f^\nu$ (in [sec]) for $\varepsilon=m_P =1.221\times 10^{22}$ MeV, T=1.3630 MeV}\label{FigNeuDecoupTimeII}\vspace{-1.25cm}
 \makebox[1\linewidth]{ \includegraphics[keepaspectratio=true,scale=0.58]{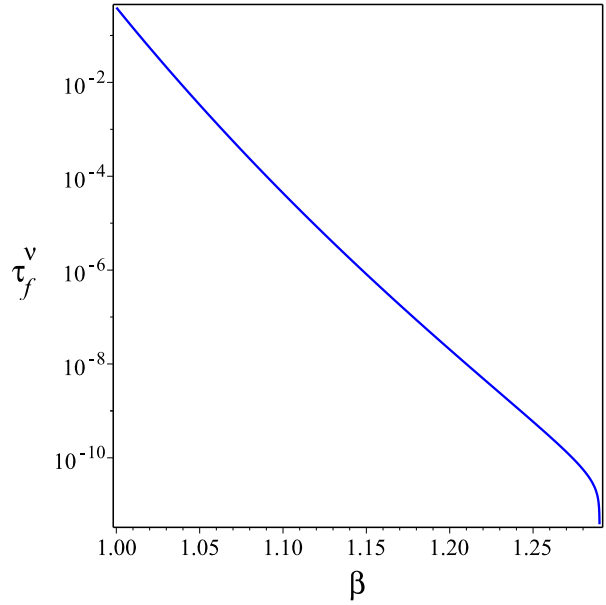}}
\end{minipage}
\vspace{-6.5cm}  \\

\end{tabular}

\begin{equation}
\begin{split}
H(x)
&=0.2887\times\sqrt{\frac{\beta}{\beta-1}}\times\left(\sqrt{\frac{(\beta-1)\,g_*}
{-5\beta^2+8\beta-2}}\right)^{1/\beta}
\left(1.1983/x^2\right)^{1/\beta} \varepsilon_{s}^{1 -1/\beta}\text{ [s}^{-1}]\\
&= H( Q )\, x^{-2/\beta}\,,
\end{split}
\end{equation}
\noindent where $H( Q )\coloneqq H(T= Q )=H(x=1)$, and $H( Q )$ is a constant carrying the parameter $\beta$.
If there were no decay of neutrons, the $X_n$ would freeze out when $\Gamma_{n\to p}(x)=H(x)$,
\begin{equation}
\frac{255}{\tau_n}\frac{x^2+6x+12}{x^5} =0.2887\times\sqrt{\frac{\beta}{\beta-1}}\times\left(\sqrt{\frac{(\beta-1)\,g_*}
{-5\beta^2+8\beta-2}}\right)^{1/\beta}
\left(1.1983/x^2\right)^{1/\beta} \varepsilon_{s}^{1 -1/\beta}\,.
\end{equation}
so the freeze-out temperature $T_{\text{MeV}}^f=\frac{1.2993}{x_f}$ can be found out by solving $x_f$ from
\begin{equation}\label{FreezeOut x}
\frac{x^2+6x+12}{x^{5-2/\beta}} =0.9963\sqrt{\frac{\beta}{\beta-1}}\times\left(\sqrt{\frac{(\beta-1)\,g_*}
{-5\beta^2+8\beta-2}}\right)^{1/\beta}
\left(1.1983\right)^{1/\beta}\varepsilon_{s}^{1 -1/\beta}\,.
\end{equation}
An exact solution to Eq.(\ref{FreezeOut x}) is difficult to work out, so it is numerically solved for a series of $\beta$ in the domain $1<\beta<(4+\sqrt{6})/5$, as shown in Tables \ref{Table I} and \ref{Table II}.\\


\subsection{Freeze-out concentration of neutrons}
\label{subsec Freeze-out concentration of neutrons}
\vspace{2mm}

To figure out the concentration $X_n$ of neutrons at the freeze-out temperature $T_f$, rewrite Eq.(\ref{Xn Nonequilibrium}) into
\begin{equation}\label{Xn Nonequilibrium II}
\begin{split}
\frac{dX_n}{dt}=\frac{dX_n}{dx}\frac{dx}{dT}\frac{dT}{dt}
=-\frac{dX_n}{dx}\cdot x \cdot\frac{\dot T}{T}
=-\Gamma_{n\to p}
\left(1+e^{-x} \right) \left(X_n -X_n^{\text{eq}}\right)\,.
\end{split}
\end{equation}
Eqs.(\ref{rho and g for radiation}) and (\ref{Hubble evolution}) imply that
\begin{equation}\label{Temperature vs time}
T=\left(\frac{30}{\pi^2 g_*}\rho\right)^{1/4}
=\left\{\frac{}{}
\frac{30\varepsilon^{2-2\beta}m_P^{2}(-5\beta^2+8\beta-2 )}{ 32\pi^3 g_*(\beta-1 )}
\Big[3\beta(\beta-1 ) \Big]^\beta\right\}^{1/4}t^{-\beta/2}
\propto t^{-\beta/2}\,,
\end{equation}
and thus ${\dot T}/{T}=-{\beta}/(2t)=-H(t)=-H(x)=-{H( Q )}{x^{-2/\beta}}$, which
recasts Eq.(\ref{Xn Nonequilibrium II}) into
\begin{equation}
\frac{dX_n}{dx}
=-\Gamma_{n\to p}\frac{x^{\frac{2}{\beta}-1}}{H( Q )}\left(1+e^{-x} \right) 
\left(X_n -X_n^{\text{eq}}\right)\,.
\end{equation}
Define a new function $F(x)\coloneqq X_n -X_n^{\text{eq}}$ to describe the departure of $X_n$ from the ideal equilibrium concentration, and transform $dX_n/dx$ into the evolution equation for $dF(x)/dx$:
\begin{equation}
\frac{dF(x)}{dx}
+\Gamma_{n\to p}\frac{x^{\frac{2}{\beta}-1}}{H( Q )}
\left(1+e^{-x} \right) F(x)
=\frac{e^x}{(1+e^x)^2}\,.
\end{equation}
Its general solution is $F(x)=\widetilde{F}(x)E(x)$, where
\begin{equation}
\widetilde{F}(x)=\exp\left[-\int^x\Gamma_{n\to p}\frac{y^{\frac{2}{\beta}-1}}{H( Q )}\left(1+e^{-y} \right)\,dy   \right]\,,
\end{equation}
and $E(x)$ satisfies
\begin{equation}
\frac{dE(x)}{dx}=\frac{1}{\widetilde{F}(x)}
\frac{e^x}{(1+e^x)^2}\,.
\end{equation}
Integrating $\widetilde{F}(x)E(x)$, we obtain
\begin{equation}
F(x)=\int^x d\tilde{x}\frac{e^{\tilde x}}{(1+e^{\tilde x})^2}
\exp\left[-\int^x_{\tilde x}\Gamma_{n\to p}\frac{y^{\frac{2}{\beta}-1}}{H( Q )}\left(1+e^{-y} \right) dy   \right]\,,
\end{equation}
and the reverse of $F(x)=X_n-X_n^{\text{eq}}$ leads to
\begin{equation}\label{Xn Nonequilibrium III}
X_n=X_n^{\text{eq}}+\int^x d\tilde{x}\frac{e^{\tilde x}}{(1+e^{\tilde x})^2}
\exp\left[-\int^x_{\tilde x}\Gamma_{n\to p}\frac{y^{\frac{2}{\beta}-1}}{H( Q )}\left(1+e^{-y} \right) dy   \right]\,.
\end{equation}
$X_n$ satisfies the initial condition $X_n(t\to 0)=X_n(T\gg Q)=X_n(x\to 0)=X_n^{\text{eq}}=1/(1+e^x)$. Without the decay of neutrons, $X_n$ would eventually freeze out after the decoupling of neutrinos; effectively setting $x=\infty$ in Eq.(\ref{Xn Nonequilibrium III}), we obtain the freeze-out concentration $X_n^\infty\coloneqq X_n(x=\infty)$:
\begin{equation}\label{Xn Nonequilibrium IV}
\begin{split}
X_n^\infty
&=\int_0^\infty d\tilde{x}\frac{e^{\tilde x}}{(1+e^{\tilde x})^2}
\exp\left[-\int^\infty_{\tilde x}\Gamma_{n\to p}\frac{y^{\frac{2}{\beta}-1}}{H( Q )}\left(1+e^{-y} \right) dy   \right]\\
&=\int_0^\infty d\tilde{x}\frac{e^{\tilde x}}{(1+e^{\tilde x})^2}
\exp\left[-\frac{255}{H( Q )\,\tau_n } \int^\infty_{\tilde x}
\left( \frac{y^2+16y+12}{y^{6-\frac{2}{\beta}}}\right)\left(1+e^{-y} \right) dy  \right]\,,
\end{split}
\end{equation}
where $X_n^{\text{eq}}(x\to\infty)=0$. Since an exact analytical result for $X_n^\infty$ is difficult (if not impossible) to find out, $X_n^\infty$ have been numerically integrated for different $\beta$, as shown in Tables \ref{Table I} and \ref{Table II}.

\begin{sidewaystable}
\begin{center}
\renewcommand\arraystretch{1.69}
\caption{$\varepsilon=1$ [1/s] or $\varepsilon_s=1$ for $H(Q)$}\label{Table I}

\begin{tabular}{|l|l|l|l|l|l|l|l|l|l|}
  \hline
  $\beta$ & $T_f^\nu$ [MeV] & $t_f^\nu$ [s] & $X_n^\nu$ & $T_f^n$ [MeV] & $t_f^n$ [s] & $H(Q)$ [1/s] & $X_n^\infty$ & $t_{\text{BBN}}$ [s] & $Y_p =2X_n^{\text{BBN}}$   \\  \hline
  1.289       & 2.4628755 & 0.026463 & 0.3716563 & 1.5171792 & 0.056119 & 8.965756 & 0.302760  & 8.638302 & 0.59952454 \\  \hline
  1.25        & 1.6377907 & 0.197342 & 0.3122414 & 0.9402799 & 0.479537 & 2.170797 & 0.218063  & 40.17368 & 0.41689076 \\  \hline
  1.2         & 1.5188824 & 0.276162 & 0.2991222 & 0.8472790 & 0.730566 & 1.662142 & 0.197121  & 61.87626 & 0.36777676 \\  \hline
  1.15        & 1.4636230 & 0.318533 & 0.2924263 & 0.7968552 & 0.917006 & 1.455879 & 0.183997  & 84.66697 & 0.33457403  \\  \hline
  1.1         & 1.4282857 & 0.344287 & 0.2879234 & 0.7585614 & 1.087931 & 1.333847 & 0.172927  & 112.8207 & 0.30461469 \\  \hline
  1.05        & 1.3997179 & 0.363572 & 0.2841493 & 0.7230330 & 1.279480 & 1.242265 & 0.161821  & 151.0530 & 0.27298984 \\  \hline
  1.01        & 1.3736317 & 0.384215 & 0.2805937 & 0.6909348 & 1.498067 & 1.166680 & 0.151417  & 195.2714 & 0.24297087  \\  \hline
  $1+10^{-3}$ & 1.3646073 & 0.393550 & 0.2793385 & 0.6814495 & 1.575961 & 1.142595 & 0.155550  & 208.7735 & 0.24575649  \\  \hline
  $1+10^{-4}$ & 1.3632380 & 0.395175 & 0.2791469 & 0.6801943 & 1.587108 & 1.139048 & 0.148162  & 210.3996 & 0.23372999 \\  \hline
  $1+10^{-5}$ & 1.3630539 & 0.395406 & 0.2791211 & 0.6800382 & 1.588536 & 1.138576 & 0.147985  & 210.5848 & 0.23339413 \\  \hline
  $1+10^{-6}$ & 1.3630308 & 0.395436 & 0.2791179 & 0.6800196 & 1.588709 & 1.138517 & 0.147967  & 210.6055 & 0.23336818 \\  \hline
  $1+10^{-7}$ & 1.3630281 & 0.395440 & 0.2791175 & 0.6800174 & 1.588730 & 1.138510 & 0.147965  & 210.6078 & 0.23336442 \\  \hline
  $1+10^{-8}$ & 1.3630277 & 0.395440 & 0.2791175 & 0.6800171 & 1.588732 & 1.138509 & 0.147965  & 210.6080 & 0.23336436  \\  \hline
  $\beta\to1^+$&1.362986  & 0.395458 & 0.2791116 & 0.6799823 & 1.588868 & 1.138509 & 0.147965 &  210.6044 & 0.23336534 \\  \hline
\end{tabular}

\end{center}
\end{sidewaystable}


\begin{sidewaystable}
\begin{center}
\renewcommand\arraystretch{1.69}
\caption{$\varepsilon=m_P$ or $\varepsilon_s=m_{ps}=0.1854\times 10^44$ for $H(Q)$}\label{Table II}

\begin{tabular}{|l|l|l|l|l|l|l|l|l|l|}
  \hline
  $\beta$ & $T_f^\nu$ [MeV] & $t_f^\nu$ [s] & $X_n^\nu$ & $T_f^n$ [MeV] & $t_f^n$ [s] & $H(Q)$ [1/s] & $X_n^\infty$ & $t_{\text{BBN}}$ [s] & $Y_p =2X_n^{\text{BBN}}$   \\  \hline
  1.289 & 1601.768 & 2.2743$\times10^{-16}$  & 0.499798&1106.8008 & 4.036$\times10^{-16}$ & 4.503 $\times 10^{10}$ & 0.49968  & 1.720$\times 10^{-9}$ &  0.999360  \\  \hline
  1.25 & 574.7002 & 3.7094$\times10^{-14}$  & 0.499437 &394.90361 & 6.761$\times10^{-14}$ & 9.778 $\times 10^{8}$ & 0.499098 & 8.919$\times 10^{-8}$ & 0.998196  \\  \hline
  1.2 & 221.2736 & 4.2086$\times10^{-12}$  & 0.498539 & 150.79103 & 7.975$\times10^{-12}$ & 2.704 $\times 10^{7}$ & 0.497615 & 3.803$\times 10^{-6}$ & 0.995230 \\  \hline
  1.15 & 78.73242 & 7.0718$\times10^{-10}$  & 0.495893 & 53.07006 & 1.404$\times10^{-9}$ & 6.409 $\times 10^{5}$ & 0.493163 & 1.923$\times 10^{-4}$ & 0.986326 \\  \hline
  1.1 & 24.60633 &2.2686$\times10^{-7}$& 0.486863 & 16.28405 & 4.805$\times10^{-7}$ &  11443.81 & 0.477623 & 0.013150 & 0.955232 \\  \hline
         1.05 & 6.481921 & 0.00017  & 0.450284 & 4.086189 & 0.00041 & 142.758 & 0.413515 & 1.31445 & 0.825796  \\  \hline
         1.01 & 1.904289 & 0.07503  & 0.336451 & 1.036945 & 0.25003 & 3.12859 & 0.226164 & 72.8185 & 0.416405  \\  \hline
  $1+10^{-3}$ & 1.410608 & 0.33343  & 0.285602 & 0.711275 & 1.30963 & 1.26217 & 0.163013 & 188.995 & 0.263015  \\  \hline
  $1+10^{-4}$ & 1.367772 & 0.38867  & 0.279780 & 0.683132 & 1.55789 & 1.15045 & 0.148929 & 208.314 & 0.235073  \\  \hline
  $1+10^{-5}$ & 1.363507 & 0.39475  & 0.279185 & 0.680332 & 1.58559 & 1.13971 & 0.148062 & 210.375 & 0.233158 \\  \hline
  $1+10^{-6}$ & 1.363076 & 0.39537  & 0.279124 & 0.680049 & 1.58841 & 1.13863 & 0.147974 & 210.585 & 0.232964  \\  \hline
  $1+10^{-7}$ & 1.363033 & 0.39543  & 0.279118 & 0.680020 & 1.58870 & 1.13852 & 0.147966 & 210.606 & 0.232946  \\  \hline
  $1+10^{-8}$ & 1.363028 & 0.39544  & 0.279117 & 0.680017 & 1.58873 & 1.13851 & 0.147965 & 210.608 & 0.232944 \\  \hline
  $\beta\to1^+$ & 1.362986 & 0.39546 &0.279112 & 0.679982 & 1.58887 & 1.13851 & 0.147965 & 210.604  &0.232943  \\  \hline
\end{tabular}

\end{center}
\end{sidewaystable}


\section{Opening of deuterium bottleneck and helium synthesis}\label{Sec Opening of deuterium bottleneck and helium synthesis}

The number densities of neutrons, protons and deuterium (D), which are nonrelativistic particles at the energy scale $T<10$ MeV, are separately
\begin{equation}
n_n=2\left(\frac{m_n T}{2\pi} \right)^{3/2}
e^{\frac{\mu_n-m_n}{T}}\,,\quad
n_p=2\left(\frac{m_P T}{2\pi} \right)^{3/2}
e^{\frac{\mu_p-m_P}{T}}\,,\quad
n_D=3\left(\frac{m_D T}{2\pi} \right)^{3/2}
e^{\frac{\mu_D-m_D}{T}}\,,
\end{equation}
so the equilibrium of chemical potentials $\mu_D=\mu_n+\mu_p$ yields
\begin{equation}\label{Deuterium concentration I}
\begin{split}
X_D\coloneqq&\,\frac{2n_D}{n_n+n_p}=\frac{3}{2}\frac{n_n n_p}{n_n+n_p}
\left(\frac{2\pi}{T}\frac{m_D}{m_n m_P}\right)^{3/2}e^{(m_n+m_P-m_D)/T}\\
=&\,\frac{3}{2}X_n X_pn_b
\left(\frac{2\pi}{T}\frac{m_D}{m_n m_P}\right)^{3/2}e^{B_D/T}\,,
\end{split}
\end{equation}
where $n_b=n_n+n_p$, and $B_D=m_n+m_p-m_D\simeq 2.2246$ MeV refers to the deuteron binding energy (with $m_D=1875.6129$ MeV, $m_n=939.5654$ MeV, and $m_P=938.2721$ MeV). Moreover, $n_b$ is related to the photon number density by
\begin{equation}\label{Deuterium nb}
n_b=\frac{g_{*s}(T)}{g_{*s}(T_0)}\times 
\eta_{10}\times 10^{-10}\times n_\gamma=\eta_{10}\times 10^{-10}\times \frac{2\zeta(3)}{\pi^2}T^3
=0.2346\times 10^{-10}\,\eta_{10} T^3\,,
\end{equation}
where $g_{*s}(T)=g_{*s}(T_0)$ after the electron-positron annihilation, and $\eta_{10}\coloneqq 10^{10}\times {n_b}/{n_\gamma}$ describes the photon-to-baryon ratio $n_b/n_\gamma$ for the net baryons left after baryogenesis.
Substituting Eq.(\ref{Deuterium nb}) into Eq.(\ref{Deuterium concentration I}), one has the deuterium concentration
\begin{equation}
\begin{split}
X_D&=10^{-10}\times \frac{3\zeta(3)}{\pi^2} \eta_{10} X_n X_p\left(\frac{2\pi}{T}\frac{m_D}{m_n m_P}\right)^{3/2}e^{B_D/T} T^3\\
&\simeq 5.6474\times 10^{-14}\times  \eta_{10} X_n X_p e^{B_D/T} T^{3/2}\,.
\end{split}
\end{equation}
Note that the value of $\eta_{10}$ can be determined through
\begin{equation}
\eta_{10}= 10^{10}\times\frac{n_b}{n_\gamma}
=10^{10}\times\frac{\rho_{\text{crit}}\Omega_b}
{m_P n_\gamma}
\simeq 274\,\Omega_b h^2 = 6.0472 \pm 0.0740\,,
\end{equation}
where $h$ denotes the Hubble constant in the unit of 100 km$\cdot$s$^{-1}$$\cdot$Mpc$^{-1}$, and we have adopted the latest Plank data $\Omega_b h^2= \Omega_b h^2 = 0.02207 \pm 0.00027$ \cite{Planck Data 2015}.
At $T_{\text{BBN}}\simeq0.079$ MeV, the $X_D$ peaks and $X_n$ drops below the concentration predicted by beta decay.
The deuterium bottleneck has broken and the remaining free neutrons 
are quickly fused into $^4$He through
 the sequence of reactions  \cite{Kolb Turner Book}
\begin{equation}
\begin{aligned}
         &n+p\to \text{D}\,,\\
      \text{D}+n&\to \text{$^3$H }+p\to\text{$^4$He}\,,\\
      \text{D}+p&\to \text{$^3$He}+n\to\text{$^4$He}\,.
\end{aligned}
\end{equation}

Following the time-temperature relation Eq.(\ref{Time-temperature relation}) with $T_{\text{MeV}}=0.079$ and $g_*\simeq 3.383538$, nucleosythesis occurs at
\begin{equation}
\begin{split}
t_{\text{BBN}}=\sqrt{3\beta(\beta-1 )}\, \left(121.5947\sqrt{\frac{ -5\beta^2+8\beta-2 }{ \beta-1}}\right)^{1/\beta} \varepsilon_s^{1/\beta -1}\text{ [s]} \,.
\end{split}
\end{equation}
Here $g_*\simeq 3.3835$; this because around the temperature $T_{\text{BBN}}\ll m_e\simeq 0.5110$ MeV after the electron-positron annihilation,
only photons and neutrinos remain as relativistic species with $T_\nu/T_\gamma= (4/11)^{1/3}$ (this ratio is independent of the number of neutrino species), hence $g_*(T\lesssim m_e)
=2+\frac{7}{8}\times3.046\times2\times\left(\frac{4}{11}\right)^{4/3}
\simeq 3.3835$. Hence, the neutron concentration at BBN is
\begin{equation}
X_n^{\text{BBN}}=X_n^\infty\,\exp
\left(\frac{t_f^n - t_{\text{BBN}}}{\tau_n}\right)\,,
\end{equation}
and the primordial $^4$He abundance is $Y_p \simeq 2X_n^{\text{BBN}}$. For different values of $\beta$, $t_{\text{BBN}}$, $X_n^{\text{BBN}}$ and $Y_p$ have been
numerically calculated, and the results have been collected in Tables \ref{Table I} and \ref{Table II}.

\section{GR limit}\label{Sec GR limit}

In the limit $\beta\to 1$, the gravitational framework reduces from $\mathscr{L}=\varepsilon^{2-2\beta}R^\beta+{16\pi}m_P^{-2}\mathscr{L}_m$ gravity to GR. In this GR limit, one has
\begin{equation}
\lim_{\beta\to 1}\sqrt{\frac{\beta}{\beta-1}}\times\left(\sqrt{\frac{\beta-1}
{-5\beta^2+8\beta-2}}\right)^{1/\beta} =1\,,
\end{equation}
and
\begin{equation}
\lim_{\beta\to 1}\sqrt{\beta(\beta-1)}\left(\frac{-5\beta^2+8\beta-2}
{\beta-1}\right)^{\frac{1}{2\beta}}=1=
\lim_{\beta\to 1}\Big[\beta(\beta-1)\Big]^\beta \left(\frac{-5\beta^2+8\beta-2}
{\beta-1}\right)\,,
\end{equation}
%
so from Eqs.(\ref{Hubble evolution}), (\ref{Time-temperature relation}) and (\ref{Temperature vs time}) one recovers the standard Hubble expansion [note: by ``standard'' we mean the standard big bang cosmology of GR]
\begin{equation}
\mathcal{H}=\frac{1}{2t}=\left(\frac{8\pi^3}{90}g_*\right)^{1/2}\frac{T^2}{m_P}
\simeq 1.6602 \sqrt{g_*}\,\frac{T^2}{m_P}
\simeq 0.2068  \sqrt{g_*}\,T^2_{\text{MeV}} \text{ [s}^{-1}]\,.
\end{equation}
as well as the the standard time-temperature relation
\begin{equation}
t=\sqrt{\frac{90}{32\pi^3}}\,g_*^{-1/2}\frac{m_P}{T^2}
\simeq \frac{2.4177}{\sqrt{g_*}\,T^2_{\text{MeV}}} \text{ [s]}
\quad\mbox{or}\quad
t T^2_{\text{MeV}}
\simeq  \frac{2.4177}{\sqrt{g_*}}\,.
\end{equation}
Equating $\mathcal{H}$ to the neutrino reaction rate $\Gamma_{\nu_e}$ in Eq.(\ref{Neutrino interaction rate}), i.e. $0.2688\, T^5_{\text{MeV}}  =0.2068 \sqrt{g_*}\,T^2_{\text{MeV}}$, one can find that neutrinos decouple at $ T=1.3630$ MeV and $t=0.3955$ [s].
Furthermore, equating $\mathcal{H}$ to the combined two-body reaction rate $\Gamma_{n\to p}$ in Eq.(\ref{Combined two-body reaction rate}),
\begin{equation}
\mathcal{H}(x)=\frac{\mathcal{H}(Q)}{x^2}=\frac{255}{\tau_n}\frac{x^2+6x+12}{x^5}\,,
\end{equation}
where $\mathcal{H}(Q)=0.3459\sqrt{g_*}$, it turns out that nucleons freeze out at
$x=1.9020$, $T_n^f=0.6800 \text{ MeV}$, and $t_n^f=1.5889$ [s].
According to Eq.(\ref{Xn Nonequilibrium III}) with $\beta\to 1$, the neutron concentration after the weak freeze-out of neutrinos is determined by
\begin{equation}
\begin{split}
X_n(x)&=X_n^{\text{eq}}+\int^x d\tilde{x}\frac{e^{\tilde x}}{(1+e^{\tilde x})^2}
\exp\left[-\int^x_{\tilde x}\Gamma_{n\to p}\frac{y}{\mathcal{H}( Q )}\left(1+e^{-y} \right) dy  \right]\\
&=X_n^{\text{eq}}+\int^x d\tilde{x}\frac{e^{\tilde x}}{(1+e^{\tilde x})^2}
\exp\left[-\frac{255}{\mathcal{H}( Q )\,\tau_n} \int^x_{\tilde x} y^{-4}\left(y^2+16y+12 \right)\left(1+e^{-y} \right) dy \right]\,,
\end{split}
\end{equation}
and thus in the absence of neutron decay $X_n$ would freeze out to the concentration  $X_n(x\to\infty)$
\begin{equation}
\begin{split}
X_n^\infty&=\int_0^\infty d\tilde{x}\frac{e^{\tilde x}}{(1+e^{\tilde x})^2}
\exp\left[- \frac{255}{\mathcal{H}( Q )\,\tau_n} \frac{\tilde{x}^2+3\tilde{x}+4+e^{-\tilde{x}}(\tilde{x}+4)}{\tilde{x}^3}  \right]=0.1480\,.
\end{split}
\end{equation}
Nucleosynthesis begins at $T\simeq0.079$ MeV, which corresponds to $t_{\text{BBN}}=210.6045$ [s]. Hence, the neutron concentration at BBN is
\begin{equation}
X_n^{\text{BBN}}=X_n^\infty\,\exp\left(\frac{t_{\text{BBN}}-t_f^n}{\tau_n}\right)=0.1167\,.
\end{equation}
and the primordial helium abundance is
\begin{equation}
Y_p \simeq 2X_n^{\text{BBN}}=0.2334.
\end{equation}
These numerical results are also collected in Tables \ref{Table I} and \ref{Table II} in the bottom row.


\section{Empirical constraints from D and $^4$He abundances}\label{Sec Empirical constraints from D and 4He abundances}

So far we have calculated the primordial nucleosynthesis in $\mathscr{L}=\varepsilon^{2-2\beta}R^\beta+{16\pi}m_P^{-2}\mathscr{L}_m$ gravity and GR from the semianalytical approach. We have seen that
primordial synthesis and abundances of the lightest elements (D, $^4$He, and also $^3$H, $^3$He, $^7$Li) rely on the baryon-to-photon ratio $\eta_{10}=10^{10}\times n_b/n_\gamma$ and the expansion rate $H$ of the Universe. In addition to the semianalytical approach, the abundances can be also be estimated in an empirical way at high accuracy \cite{BBN empirical I, BBN empirical II}. For nonstandard expansion $H=\frac{\beta}{2t}$ in $\mathscr{L}=\varepsilon^{2-2\beta}R^\beta+{16\pi}m_P^{-2}\mathscr{L}_m$ gravity that deviates from the standard expansion $\mathcal{H}=\frac{1}{2t}$ in GR, employ the nonstandard-expansion parameter
\begin{equation}
S\coloneqq\frac{H}{\mathcal{H}}\quad\Rightarrow\quad S=\beta\,.
\end{equation}
It has been found that, for the priors $4\lesssim \eta_{10}\lesssim8$ and $0.85\lesssim S\lesssim1.15$, the primordial deuterium and $^4$He abundances satisfy the best-fit formulae \begin{equation}\label{Empirical D}
y_{\text D}\coloneqq 10^5 \times  \frac{\text D}{^1\text H}
= 46.5\times(1\pm 0.03)\times \Big[\eta_{10}-6(S-1)\Big]^{-1.6}
\end{equation}
and
\begin{equation}\label{Empirical He4}
Y_p=(0.2386\pm 0.0006)+2\times10^{-4}\times(\tau_n-885.7)
+\frac{\eta_{10}}{625}+\frac{S-1}{6.25}\,,
\end{equation}
the reverse of which respectively yield
\begin{equation}\label{Empirical D S}
S =\frac{\eta_{10}}{6} - \frac{1}{6}
\left[\frac{46.5\times(1\pm 0.03)}{y_{\text D}}\right]^{1/1.6} +1
\end{equation}
and
\begin{equation}\label{Empirical He4 S}
S= 6.25 \times \Big[ Y_p-(0.2386\pm 0.0006) +2\times10^{-4}\times(885.7-\tau_n)\Big]-\frac{\eta_{10}}{100}+1\,.
\end{equation}
Recall that $\eta_{10}= 6.0472 \pm 0.0740$ for $\Omega_b h^2 = 0.02207 \pm 0.00027$, $\tau_n=880.0\pm 0.9$ [s], and according to the recommended values from the Particle Data Group \cite{BBN Particle Data Group}, we have
\begin{equation}
y_{\text D}=2.53 \pm 0.04 \quad,\quad
Y_p=0.2465 \pm 0.0097\,.
\end{equation}
Thus, Eqs.(\ref{Empirical D S}) and (\ref{Empirical He4 S}) lead to
\begin{equation}\label{Empirical D beta}
S=0.9797\pm 0.0708
\quad\text{or}\quad 0.9089\leq S=\beta \leq 1.0505 \quad\text{(deuterium)}\,,
\end{equation}
\begin{equation}\label{Empirical He4 beta}
S=0.9960\pm   0.1036 \quad\text{or}\quad
\quad 0.8925\leq S=\beta \leq 1.0996  \quad\text{($^4$He)}\,.
\end{equation}
Here for the errors of mutually independent quantities in $\{x_i\pm \Delta x_i, x_j\pm \Delta x_j \}\mapsto y+\Delta y$, we have applied the propagation rules that $\Delta y=\sqrt{\left(\Delta x_i  \right)^2+\left(\Delta x_j  \right)^2 }$ for $y=x_i \pm x_j$, $
       \frac{\Delta y}{y}=\sqrt{\left(\frac{\Delta x_i}{x_i}\right)^2+ \left(\frac{\Delta x_j}{x_j}\right)^2 }$ for $y=x_i x_j$ or $y=x_i/x_j(i\neq j)$, and $\Delta y= \sqrt{n}\,x^{n-1}\Delta x$ for $y=x^n$.

Combining Eq.(\ref{Empirical D beta}) with Eq.(\ref{Empirical He4 beta}), we find $0.9089\leq S=\beta \leq 1.0505$; taking into account the positive energy density/temperature condition $1<\beta<(4+\sqrt{6})/5$ in Eq.(\ref{beta domain I}), we further obtain $1< S=\beta \leq 1.0505$. Since $S$ is related to the extra number of effective neutrino species by
\begin{equation}
S\coloneqq\frac{H}{\mathcal{H}}=\left(1+\frac{7}{43}
\Delta N_\nu \right)^{1/2}
\quad\Rightarrow\quad
\Delta N_\nu=\frac{43}{7}
(\beta^2-1)\,,
\end{equation}
thus for $1< S=\beta \leq 1.0505$, $\Delta N_\nu^{\text{eff}}\coloneqq N_\nu^{\text{eff}}-3$ is constrained by
\begin{equation}
0< \Delta N_\nu^{\text{eff}} \leq 0.6365\,.
\end{equation}

Note that the theoretically predicted primordial abundance for $^7$Li is found to respect the best-fit formula
\begin{equation}
y_{\text{Li}}\coloneqq 10^{10}\times \frac{\text{Li}}{^1\text H} 
= \frac{(1\pm 0.1)}{8.5}\times \Big[\eta_{10}-3(S-1) \Big]^2\,,
\end{equation}
which, for the domain $1 < S=\beta \leq 1.0505$, gives rise to
\begin{equation}
y_{\text{Li}}=4.0892\pm 0.0012\; (\beta=1.0505)\quad\text{to}\quad
4.3022\pm 0.0012 \; (\beta=1)\,.
\end{equation}
Hence,
\begin{equation}
4.0880\leq y_{\text{Li}}< 4.3034\,,
\end{equation}
which is much greater than the observed abundance $y_{\text{Li}}=1.6 \pm 0.3$ \cite{BBN Particle Data Group}. This indicates that the lithium problem remains unsolved in $\mathscr{L}=\varepsilon^{2-2\beta}R^\beta+{16\pi}m_P^{-2}\mathscr{L}_m$ gravity.



\section{Consistency with gravitational baryogenesis}\label{Sec Consistency with gravitational baryogenesis}

We just investigated the primordial nucleosynthesis in $\mathscr{L}=\varepsilon^{2-2\beta}R^\beta+16\pi m_P^{-2}\mathscr{L}_m$ gravity from the semianalytical and the empirical approaches. The nucleons building the lightest nuclei come from the net baryons left after baryogenesis, and in this section we will quickly check the consistency of $\mathscr{L}=\varepsilon^{2-2\beta}R^\beta+16\pi m_P^{-2}\mathscr{L}_m$ gravity with the baryon-antibaryon asymmetry using the framework of gravitational baryogenesis \cite{Davoudiasl Gravitational Baryogenesis}, which,  compared with traditional Sakharov-type mechanisms,
dynamically produces the required baryon asymmetry for an expanding Universe by violating the combined symmetry of charge conjugation, parity transformation and time reversal (CPT) while being in thermal equilibrium.
In this mechanism, the dominance of baryons over antibaryons attribute to the coupling between the gradient of the Ricci curvature scalar $R$ and some current $J^\mu_B$ leading to net baryon-lepton charges:
\begin{equation}
\int d^4x\sqrt{-g}\, \frac{\left(\partial_\mu R\right)J^\mu_B}{M_*^2}
=\int d^4x\sqrt{-g}\,\frac{\dot{R}\left(n_B-n_{\bar B}\right)}{M_*^2}\,,
\end{equation}
where $M_*$ refers to the cutoff scale of the effective theory, and is estimated to take the value of the reduced Plank mass $M_*\simeq m_P/\sqrt{8\pi}$.

The baryon asymmetry can be depicted by the dimensionless baryon-to-entropy ratio $n_B/s$ of the radiation-dominated Universe, with
\begin{equation}\label{Baryogenesis I}
n_B\simeq\frac{1}{6}g_b \mu_B T^2\quad\mbox{and}\quad
s=\frac{2\pi^2}{45}g_{*s}T^3\,,
\end{equation}
where $g_b=28=2 \text{ (photon)}+2\times 8 \text{ (gluon)}+3\times 3\, (W^{\pm}, Z^0)+1 \text{ (Higgs})$ for $T>m\text{(top quark)}\simeq 1.733 \times 10^{5}$ MeV, and $\mu_B\coloneqq-{\dot{R}}/{M_*^2}$ acts as the effective chemical potential. Also, $g_{*s}$ denotes the entropic effective \\

\begin{tabular}{c}

 \begin{minipage}{\linewidth}
\captionof{figure}{${n_B}/{s}$ for $\varepsilon=1 \text{ sec}^{-1}=6.58\times 10^{-22}$ MeV}\label{FigBaryogenesisI}\vspace{-1.25cm}
 \makebox[1\linewidth]{ \includegraphics[keepaspectratio=true,scale=0.6]{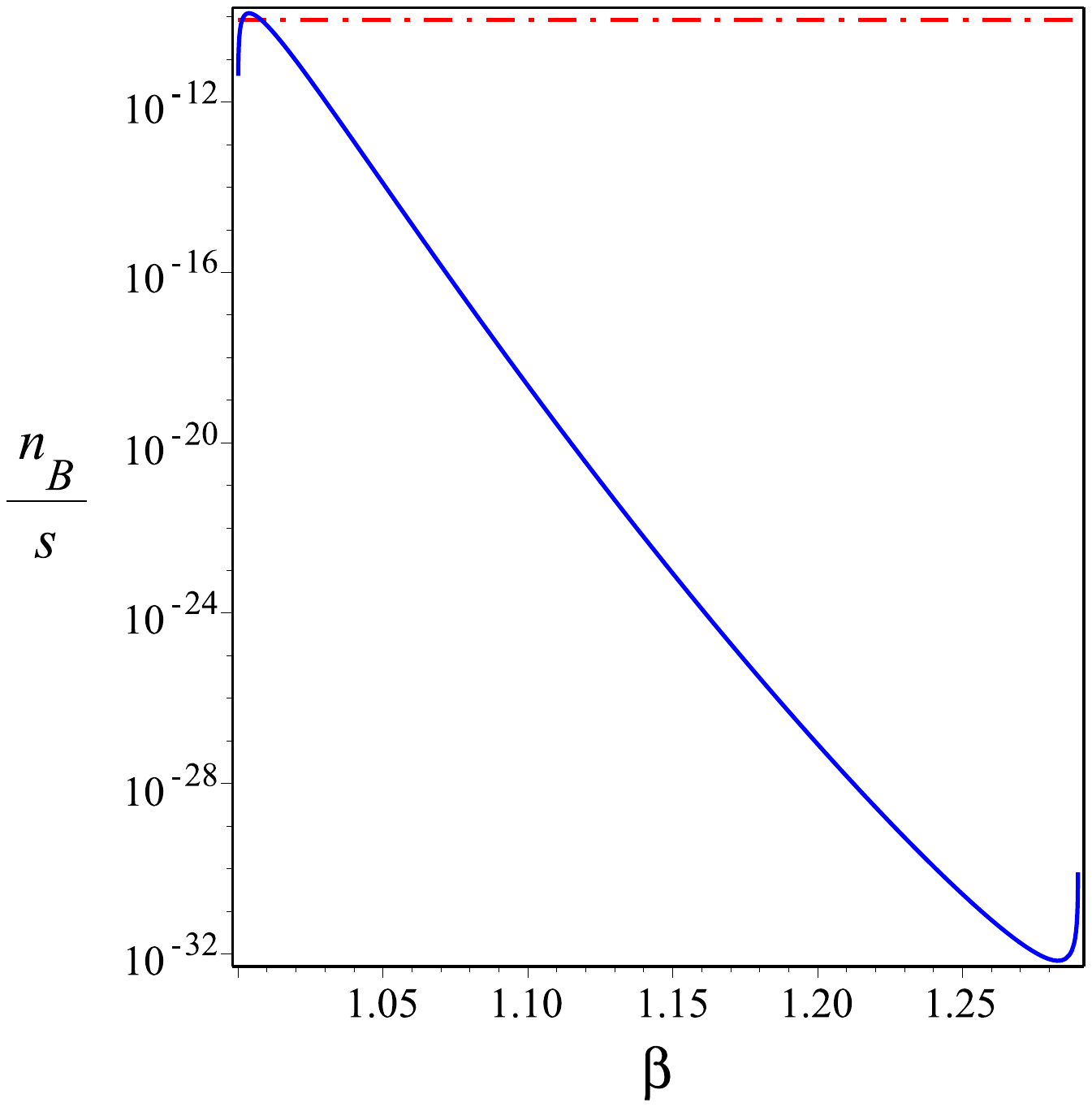}}
\end{minipage}
\vspace{-6.5cm}\\

  \begin{minipage}{\linewidth}
\captionof{figure}{${n_B}/{s}$ for $\varepsilon=1 \text{ sec}^{-1}=6.58\times 10^{-22}$ MeV}\label{FigBaryogenesisII}\vspace{-1.25cm}
 \makebox[1\linewidth]{ \includegraphics[keepaspectratio=true,scale=0.65]{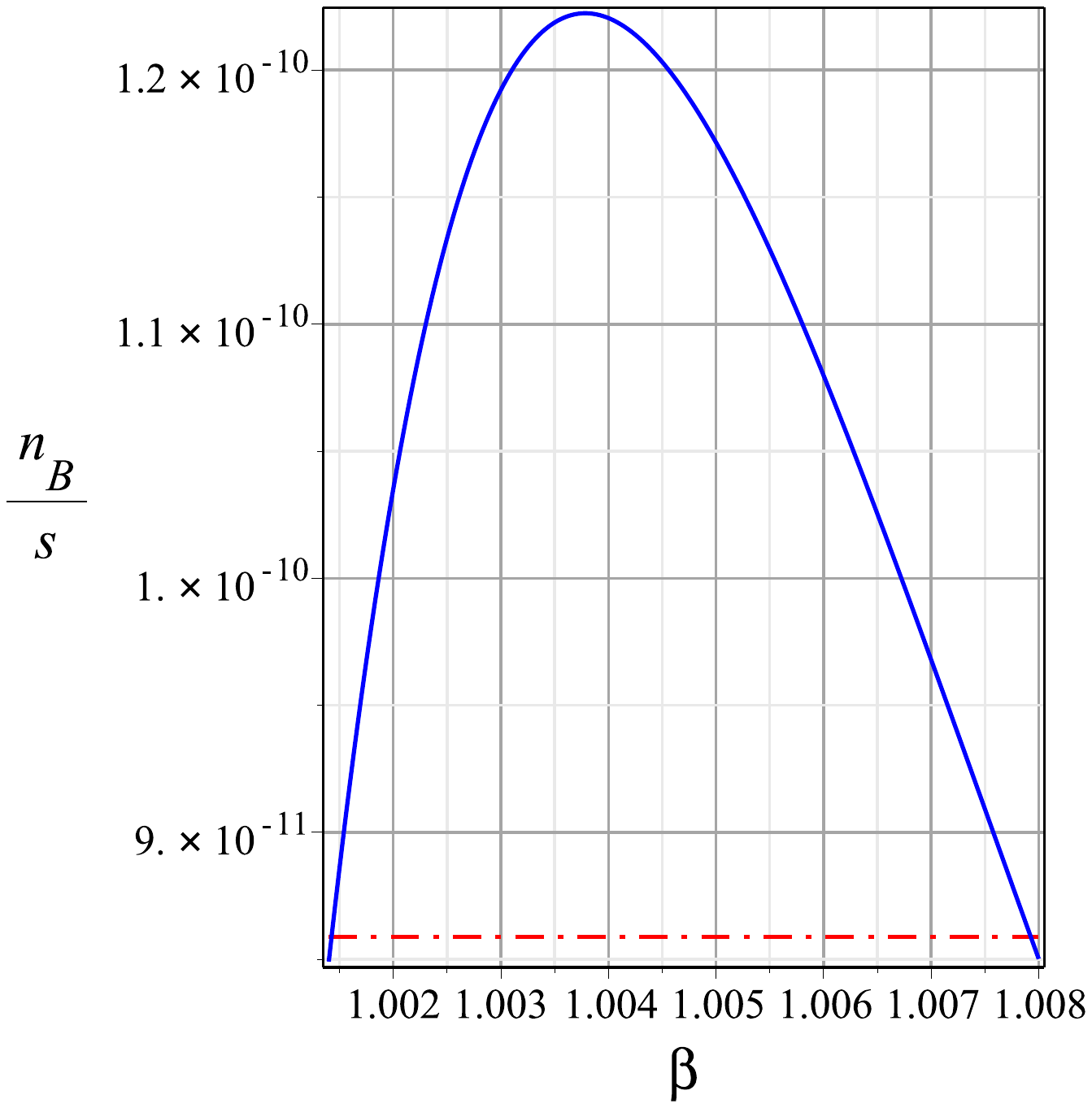}}
\end{minipage}
\vspace{-6.5cm}\\

\end{tabular}

  \begin{minipage}{\linewidth}
\captionof{figure}{${n_B}/{s}$ for $\varepsilon=m_P$}\label{FigBaryogenesisIII}\vspace{-1.25cm}
 \makebox[1\linewidth]{ \includegraphics[keepaspectratio=true,scale=0.6]{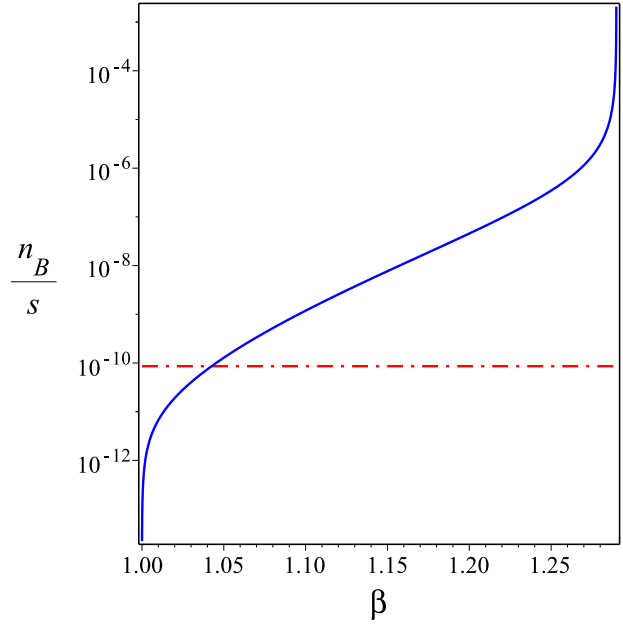}}
\end{minipage}
\vspace{-6.5cm}

\noindent number of degree of freedom, and is defined like $g_*$ by
\begin{equation}
g_{*s}\coloneqq\sum_{\text{boson}} g_i^{(b)}\left(\frac{T_{i}}{T}\right)^3+\frac{7}{8}\sum_\text{fermion} g_j^{(f)} \left(\frac{T_{j}}{T}\right)^3\,;
\end{equation}
\noindent one has
$g_{*s}=g_*$ at the baryogenesis era when all standard-model particles are relativistic and in equilibrium, $g_f=2\times 3 \text{ (neutrino)}+2\times 6 \text{ (charged lepton)}+12\times 6 \text{ (quark)}=90$\footnote{When calculating $g_f$, we use $N_\nu^{\text{eff}}=3$ rather than $N_\nu^{\text{eff}}=3.046$, because baryogenesis happens before primordial nucleosythesis and it's unnecessary to consider the ``non-instantaneity'' of neutrinos' decoupling.},
and $g_{*s}=g_*=g_b+\frac{7}{8}g_f=106.75$.
In $\mathscr{L}=\varepsilon^{2-2\beta}R^\beta+16\pi m_P^{-2}\mathscr{L}_m$ gravity for which $\partial_\mu R$ or $\dot R$ is nontrivial, Eqs.(\ref{Time-temperature relation}) and (\ref{Baryogenesis I}) lead to
\begin{equation}\label{Baryon to entropy ratio}
\begin{split}
\frac{n_B}{s}
&=\left.-\frac{15}{4\pi^2}\frac{g_b}{g_{*s}}\frac{\dot{R}}{M_*^2 T}\right|_{T_d}
=\frac{45}{2\pi^2}\frac{g_b}{g_{*s}}\frac{\beta(\beta-1)}{t^3 M_*^2 T_d}\\
&=\frac{5\sqrt{3}}{2\pi^2}\frac{g_b}{g_{*s}}\frac{1}{\sqrt{\beta(\beta-1 )}}
\left(\sqrt{\frac{ (\beta-1)\,g_*}{ -5\beta^2+8\beta-2 }}\right)^{3/\beta}
\left(\sqrt{\frac{32\pi^3}{30}}\frac{T_d^2}{\varepsilon m_P}\right)^{3/\beta}\frac{\varepsilon^3}{M_*^2 T_d} \,,
\end{split}
\end{equation}
where $T_d\simeq 3.3\times 10^{19} \text{ MeV}$ is the upper bound on the tensor-mode fluctuations at the inflationary scale \cite{Lambiase and Scarpetta Gravitational Baryogenesis fR}.


Following the observational value $ \Omega_b h^2 = 0.02207 \pm 0.00027$ \cite{Planck Data 2015}, we have the net-baryon-to-entropy ratio $n_b/s=\frac{n_b}{n_\gamma}/7.04=3.8920\times 10^{-9}\Omega_b h^2=(8.5897\pm 0.1051)\times 10^{-11}$, which remains constant during the expansion of the early Universe and imposes a constraint to $n_B/s$. For $\varepsilon=\text{ [s}^{-1}]=6.58\times 10^{-22} \text{ MeV}$, Eq.(\ref{Baryon to entropy ratio}) respects this constraint for all $1<\beta<(4+\sqrt{6})/5$, as shown in Figure \ref{FigBaryogenesisI}, with minor violation for $1.001426<\beta< 1.007925$, as magnified in Fig. \ref{FigBaryogenesisII}; however, this minor violation can be easily removed by a fluctuation of $M_*$ and $T_d$. For $\varepsilon=m_P$, this constraints is satisfied for $1<\beta<1.04255$, as shown in Fig. \ref{FigBaryogenesisIII}.

\section{Conclusions}


In this paper, we have investigated the nonstandard BBN in  $\mathscr{L}=\varepsilon^{2-2\beta}R^\beta+{16\pi}m_P^{-2}\mathscr{L}_m$ gravity. The main results,
compared with the standard BBN or the GR limit in Sec.~\ref{Sec GR limit},
include Eq.(\ref{Hubble evolution}) for the nonstandard Hubble expansion, Eq.(\ref{Time-temperature relation}) for the generalized time-temperature correspondence, Eq.(\ref{Neutrino decoupling temperature}) for the neutrino decoupling temperature $T_f^\nu$, Eq.(\ref{FreezeOut x}) for the freeze-out temperature $T_f^n$ of nucleons, Eq.(\ref{Xn Nonequilibrium III}) for the out-of-equilibrium concentration $X_n$, Eq.(\ref{Xn Nonequilibrium IV}) for the freeze-out concentration $X_n^\infty$. From the data points in Tables \ref{Table I} and \ref{Table II}, we have shown that every step of BBN is considerably $\beta-$dependent when running over the entire domain $1<\beta<(4+\sqrt{6})/5$.

In the semianalytical approach, $\beta$ is constrained to $1<\beta<1.05$ for $\varepsilon=1$ [1/s] and $1<\beta<1.001$ for $\varepsilon=m_P$. In the empirical approach, we have found $1< \beta \leq 1.0505$ which corresponds to an extra number of neutrino species by $0< \Delta N_\nu^{\text{eff}} \leq 0.6365$.
In theory, it might be possible for modified gravities
to severely rescale the thermal history of the early Universe without changing the state of the current Universe. This requires the joint investigations of BBN, the cosmic radiation background, and the structure formation, and we will look into the possibility of such strongly modified gravities in our prospective studies.


\section*{Acknowledgement}

This work  was supported by NSERC grant 261429-2013.

\end{document}